\documentclass[conference, 10pt]{IEEEtran}
\usepackage{graphics}
\usepackage{amsmath,amsfonts,amssymb,varioref,mathrsfs,verbatim}
\usepackage{epsfig}
\usepackage{array}
\usepackage[english]{babel}
\usepackage{color}
\usepackage{babel}

\usepackage{epstopdf}
\usepackage{color}
\renewcommand{\figurename}{Fig.}
\renewcommand{\figurename}{Fig.}
\addto\captionsenglish{\renewcommand{\figurename}{Fig.}}
\begin{document}
\pagenumbering{gobble}

\title{Millimeter-Wave Antenna Array Diagnosis with Partial Channel State Information}
\author{George Medina$^*$, Akashdeep Singh Jida$^*$, Sravan Pulipati$^{\ddag}$, Rohith Talwar$^*$, Nancy Amala J$^*$,\\  Tareq Y. Al-Naffouri$^{\S}$, Arjuna Madanayake$^{\ddag}$ and Mohammed E. Eltayeb$^*$ \\ \small
$^*$Department of Electrical \& Electronic Engineering  California State University, Sacramento, USA. Email: mohammed.eltayeb@csus.edu\\
$^{\ddag}$Electrical and Computer Engineering, Florida International University, Miami, USA. Email: amadanay@fiu.edu\\
$^{\S}$King Abdullah University of Science and Technology, Thuwal, KSA. Email: tareq.alnaffouri@kaust.edu.sa
}
\maketitle

\maketitle

\begin{abstract}
Large antenna arrays enable directional precoding for Millimeter-Wave (mmWave) systems and provide sufficient link budget to combat the high path-loss at these frequencies. Due to atmospheric conditions and hardware malfunction, outdoor mmWave antenna arrays are prone to blockages or complete failures. This results in a modified array geometry, distorted far-field radiation pattern, and system performance degradation. Recent remote array diagnostic techniques have emerged as an effective way to detect defective antenna elements in an array with few diagnostic measurements. These techniques, however, require full and perfect channel state information (CSI), which can be challenging to acquire in the presence of antenna faults. This paper proposes a new remote array diagnosis technique that relaxes the need for full CSI and only requires knowledge of the incident angle-of-arrivals, i.e. partial channel knowledge. Numerical results demonstrate the effectiveness of the proposed technique and show that fault detection can be obtained with comparable number of diagnostic measurements required by diagnostic techniques based on full channel knowledge. In presence of channel estimation errors, the proposed technique is shown to out-perform recently proposed array diagnostic techniques.
\end{abstract}

\begin{IEEEkeywords}
Antenna Arrays, fault diagnosis, compressed sensing, millimeter-wave communication.
\end{IEEEkeywords}

\IEEEpeerreviewmaketitle

\section{Introduction}
Communication in the Millimeter-Wave (mmWave) band is the new frontier for next generation wireless systems \cite{mag}-\cite{I3}. To provide sufficient link budget for these system, large antenna arrays will be required to enable directional precoding \cite{I3}-\cite{Sarieddeen2019}. Thanks to the small carrier wavelength at mmWave frequencies, multiple antenna elements can be packaged onto a small chip, possibly with other RF components \cite{repa}. The densely packed antenna elements, however, introduces new challenges for these systems. The comparable sizes of blockages, e.g., dirt, water droplets, and ice, can completely (or partially) block mmWave signals incident on a single or multiple antenna elements. Manufacturing imperfections can also lead to antenna element failure. Antenna element blockage or failure randomizes the array's geometry, and as a result, distorts its radiation pattern and causes uncertainties in the mmWave channel \cite{m0}.  Therefore, it is crucial to design remote array diagnosis techniques that continuously monitor the performance of mmWave antenna arrays and minimize the effects of antenna element failures.

Several techniques based on sparse signal recovery have recently emerged in the literature to remotely diagnose antenna arrays in a fast and reliable manner \cite{m0}-\cite{de}. These techniques formulate the fault detection problem as a sparse signal recovery problem in compressed sensing. Specifically, a sparse \textit{difference response} vector is generated by subtracting the response of a reference fault-free antenna array from the response of a potentially faulty antenna array, commonly known as the \textit{Array-Under-Test} (AUT). Using this sparse difference response vector, sparse signal recovery algorithms, see e.g. \cite{cs1}-\cite{cs4},  are then applied to recover identity of the faulty antenna elements.  We refer to such techniques  as \textit{difference based} techniques in this paper. Other techniques adopt a deep learning approach to diagnose mmWave antenna arrays \cite{ml1} \cite{ml2}. These techniques apply machine learning algorithms to identify faulty antenna elements by measuring distortions in the far-field radiation pattern. Despite their excellent performance, the above referenced diagnosis techniques require full and perfect channel state information (CSI) to generate and update the response of the reference fault-free antenna array in a timely manner. This is challenging since perfect CSI estimation is dependent on many factors e.g. link quality, number of scatters, estimation errors, etc., and the faulty array itself distorts the communication channel estimate \cite{d5}. To overcome this limitation, it is curial that new array diagnosis techniques are design to be independent on prior communication channel knowledge.

In this paper, we propose a new technique for remote array diagnosis. The proposed technique only requires knowledge of the set of all possible \textit{Angle-of-Arrivals} (AoAs) the diagnostic signals take, and does not require full channel knowledge. The idea is to design the combining vector (or antenna weights) at the AUT to null diagnostic signals from all incident AoAs. In the presence of antenna faults, the receive beam pattern is distorted, and diagnostic signals can not be nulled. These received (or leaked) diagnostic signals are exploited to formulate the diagnosis problem as a sparse signal recovery in compressed sensing. As we will show in Section III, this technique enables antenna fault detection with just a few diagnostic measurements.  The main contributions of this paper can be summarized as follows: (i) We present new array diagnosis formulation that takes the effect of the communication channel into account. Prior work assumes perfect knowledge of the far-field beam pattern and does not take the effect of the communication channel into account. (ii)  We present a new and novel array diagnosis technique that relaxes the need for full channel knowledge. To the best of our knowledge, this is the first paper that proposes an array diagnosis technique that requires partial channel knowledge.

The remainder of this paper is organized as follows. In Section II, we formulate the mmWave antenna array diagnosis problem in the presence of multipath. In Section III, we present the proposed array diagnosis technique. In Section IV, we present some numerical results and conclude our work in Section V.

\section{Problem Formulation}
We consider a transceiver equipped with a uniform linear antenna array which consists of $N$ equally spaced elements and $S << N$ possibly faulty elements. A fault is defined as any impairment that causes an antenna element to function abnormally.  A fault can result from either physical blockage of an antenna element or circuit failure.  While a linear array is adopted in this paper for simplicity, other antenna structures can be equally adopted. 

To initiate antenna diagnosis, a probe is used to transmit $M$ diagnosis symbols to the transceiver with the AUT. In the absence of antenna faults, the $m$th received diagnosis symbol can be written as
\begin{eqnarray} \label{y1}
 y_m = \mathbf{w}_m^*\mathbf{h} s + z_m, 
\end{eqnarray}
where the entries of the combining vector $\mathbf{w}_m\in\mathcal{C}^{N\times 1}$ represent the $m$th complex weights associated with the receive antenna, $\mathbf{h}$ is the mmWave channel between the transceiver and the probe, $s_m$ is the $m$th transmitted diagnosis symbol, and $z_m \sim \mathcal{CN}(0,\sigma^2)$ is the additive noise at the transceiver. A geometric channel model with $L$ scatterers is adopted in this paper \cite{I3} \cite{ahmed} \cite{rap}.  Under this model, the channel can be expressed as
	\begin{eqnarray}\label{channelk}
	\mathbf{h} = \sqrt{\frac{N}  {L}} \sum_{\ell=1}^L \alpha_{\ell} {\mathbf{a}}(\theta_{\ell}),
	\end{eqnarray}
where $\alpha_{\ell} \sim \mathcal{CN} (0,1)$ is the complex gain of the $\ell$th path, $\theta_{\ell}$ is the $\ell$th path AoA, and the vector  ${\mathbf{a}}(\theta_{\ell})$  represents the transceiver's antenna array response corresponding to the $\ell$th AoA $\theta_{\ell}$.  For simplicity, we set $s=1$ in (\ref{y1}) and omit it from the subsequent analysis. 

In the presence of antenna faults, the received diagnosis symbol becomes
\begin{eqnarray} \label{y2}
 \hat{y}_m &=& \mathbf{w}_m^*{\mathbf{Bh}} + z_m \\  \label{y3}
  &=& \mathbf{w}_m^*\hat{\mathbf{h}} + z_m
\end{eqnarray}
where  $\hat{\mathbf{h}} = {\mathbf{Bh}} $ is the equivalent mmWave channel. The entries of the diagonal matrix $\mathbf{B} \in \mathcal{C}^{N\times N}$ are
  \begin{equation}\label{efbp1}
\text{B}_{n,n}  = \left\{
               \begin{array}{ll}
               \alpha_n, & \hbox{ if the $n$th antenna element is faulty}  \\
               1, & \hbox{ otherwise,  } \\
               \end{array}
               \right.
\end{equation}
where $\alpha_n = \kappa_n e^{j\Phi_n}$,  $0 \leq \kappa_{n} \le 1$ and $0 \leq \Phi_{n} \leq 2\pi$.  A value of $\kappa_{n} = 0$ represents maximum blockage (or complete failure), and $\Phi_{n}$ captures the phase-shift caused by the fault at the $n$th antenna element.  The diagonal matrix $\mathbf{B}$ captures failures that can result from internal circuitry of the antenna element itself, or from external blockages caused by, for example, weather.  From equations (\ref{y2}) and (\ref{y3}), we observe that faults modify the antenna array manifold and causes uncertainty the mmWave channel.

To locate the identity of the faulty antenna elements, prior work in the literature proposed several techniques which are based on subtracting $M$ received diagnosis symbols in (\ref{y3}) from $M$ ideal (fault-free) diagnosis symbols in (\ref{y1}). The ideal diagnosis symbols in (\ref{y1}) can be generated offline if the channel $\mathbf{h}$ is fully known at the receiver. This subtraction results in  the following difference vector $\mathbf{y}_\text{d} \in \mathcal{C}^{M\times 1}$
\begin{eqnarray} \label{yd}
 \mathbf{y}_\text{d} &=&    \mathbf{y} -  \hat{\mathbf{y}}  \\  \label{yd2}
   &=& \mathbf{W}^*{\mathbf{h}}-\mathbf{W}^*\hat{\mathbf{h}} + \mathbf{z} \\  \label{yd3}
      &=& \mathbf{W}^*{\mathbf{h}}_\text{d} + \mathbf{z}.
\end{eqnarray}
In (\ref{yd2}) and (\ref{yd3}), the matrix $\mathbf{W} = [\mathbf{w}_1, \mathbf{w}_2, ...., \mathbf{w}_M]$, $ \mathbf{z}$ is the additive noise vector, and the  vector ${\mathbf{h}}_\text{d}$ is sparse with the non-zero entries corresponding to the identity of the faulty antenna elements.  Applying any sparse recovery techniques, see e.g. \cite{cs1}-\cite{cs4}, ${\mathbf{h}}_\text{d}$ can be recovered with overwhelming probability from $ \mathbf{y}_\text{d}$ and $\mathbf{W}$ provided that the number of diagnosis symbols $M> 2S \log{N}$. 

While excellent, the requirement of perfect channel knowledge is not practical and poses a greater challenge for difference based techniques. Perfect channel knowledge might not be readily available in practice, and as shown in \cite{d5}, acquiring perfect channel estimation is not possible with faulty antenna hardware. 
In the following section, we propose a new approach to identify the faulty antenna elements. The proposed approach relaxes the need for full channel knowledge and only requires a set of all possible angles of arrivals. The receive angles of arrival can be easily obtained by, for example, beam training techniques outlined in \cite{bt1}-\cite{ bt4} and references therein, or provided by finger-printing techniques outlined in \cite{F1}-\cite{F3} and references therein.

\section{Antenna Fault Detection at the Receiver}
In this section we mathematically formulate and outline the proposed antenna fault detection technique. For this technique, we assume that the receiver is only equipped with knowledge of its angles of arrivals $\theta_{\ell} \in \Theta$, where $\Theta$ is a set that contains all possible AoAs. The receiver has no knowledge of the complex paths gains nor their corresponding delays (if any). To mathematically formulate the problem solution, we first rewrite (\ref{y3}) as
 \begin{eqnarray} \label{yp1}
 \hat{y}_m = \mathbf{w}_m^*(\mathbf{h} + \mathbf{h}_\text{e})  + z_m,
\end{eqnarray}
where $\mathbf{h}_\text{e}$ is the error in the mmWave channel caused by faulty receive antennas. Observe that $\mathbf{h}_\text{e}$ is sparse with the non-zero elements corresponding to the fault locations. If the AoAs are quantized to $N$ points,  the channel in (\ref{yp1}) can be expressed in matrix form as
 \begin{eqnarray} \label{yp2}
 \hat{y}_m &=& \mathbf{w}_m^*(\mathbf{A} + \mathbf{A}_\text{e})\mathbf{x}  + z_m\\ \label{yp21}
 &=& \mathbf{w}_m^*\mathbf{Ax} + \mathbf{w}_m^*\mathbf{A}_\text{e}\mathbf{x}  + z_m,
\end{eqnarray}
 where the matrix $\mathbf{A}$ is the DFT matrix with its $i$th column corresponding to the array response of $i$th quantized AoA. The $L$ non-zero entries of the sparse vector $\mathbf{x}$ correspond to the complex gains of the $L$ paths. The matrix $\mathbf{A}_\text{e}$ is row sparse with its non-zero row entries corresponding to the error imposed by the faulty antenna elements. 
 
 As the objective of this paper is to detect antenna faults (the second term in (\ref{yp21})), it is imperative that the weights $\mathbf{w}_m$ are designed to be in the null-space of the column vectors of the DFT matrix $\mathbf{A}$ that correspond to the $L$ AoAs. There are two main ways to achieve this. If the AoAs are quantized to $N$ points, then one can exploit the orthogonality property of the DFT matrix $\mathbf{A}$ in (\ref{yp2}) to select the columns that do not correspond to the $L$ AoAs as the receive beamforming (or measurement) weights, i.e. $\mathbf{w}_m \in [\mathbf{A}]_{:,m}, m \not= l$.  If the AoAs are not quantized, the vector $\mathbf{w}_m$ needs to be orthogonal to all AoAs.  Exploiting the large antenna dimensions available in mmWave systems, one can generate $M$ receive antenna weights (or beam vectors) that are orthogonal to the array response corresponding to directions in $\Theta$. To achieve this, let the matrix $\mathbf{D} = [{\mathbf{a}}(\theta_{1}), {\mathbf{a}}(\theta_{2}),..., {\mathbf{a}}(\theta_{L})]$ contain the array response vectors that correspond to the $L$ AoAs in $\Theta$. Using Householder transformation \cite{householder}, the orthogonal beam matrix $\mathbf{Q} \in \mathcal{C}^{N\times N}$ can be obtained as follows
\begin{eqnarray} \label{HH}
 \mathbf{Q} =    \mathbf{I} -  \mathbf{D}(\mathbf{D}^*\mathbf{D})^{-1}\mathbf{D}^*.
\end{eqnarray}
The combining matrix ${\mathbf{W}}$ is then formed by selecting $M$ columns from the matrix $\mathbf{Q}$. 

 To this end note that the combining matrix $\mathbf{W}$ is used to receive the $M$ diagnosis symbols as follows
  \begin{eqnarray} \label{yp3}
 \hat{\mathbf{y}}  &=& \mathbf{W}^*\mathbf{Ax} + \mathbf{W}^*\mathbf{A}_\text{e}\mathbf{x}  + \mathbf{z}\\ \label{yp31}
  &=& \mathbf{W}^*\mathbf{A}_\text{e}\mathbf{x}  + \tilde{\mathbf{z}}+ \mathbf{z}\\ \label{yp32}
    &=& \mathbf{W}^*\mathbf{h}_\text{e} + \tilde{\mathbf{z}} + \mathbf{z}.
\end{eqnarray}
 Note as the columns of $\mathbf{W}$ are orthogonal to the columns in $\mathbf{A}$ corresponding to the $L$ AoAs,  the first term in (\ref{yp3}) cancels out. The interference vector $\tilde{\mathbf{z}}$  accounts for the interference that arises when  $\mathbf{W}^*$ and $\mathbf{Ax}$ are not orthogonal. This situation could arise due to imperfect channel estimates at the receiver. As the error vector $\mathbf{h}_e$ is sparse, with non-zero entries corresponding to the faulty antenna elements, compressed sensing techniques outlined in \cite{cs1}-\cite{cs4} can be used to recover $\mathbf{h}_\text{e} $ from $ \hat{\mathbf{y}}$ and $\mathbf{W}$ as follows:
  \begin{eqnarray} 
  \nonumber \min   && ||\tilde{\mathbf{h}}_\text{e} ||_1 \\
 \nonumber \text{s.t.}    &&|| \hat{\mathbf{y}}  - \mathbf{W}^*\tilde{\mathbf{h}}_\text{e} ||_2 \leq \epsilon.
\end{eqnarray}
For simplicity, we employ the orthogonal matching pursuit (OMP) algorithm \cite{cs4} to solve the above optimization problem. The non-zero entries of the recovered vector $\tilde{\mathbf{h}}_\text{e} \in \mathcal{C}^{N \times 1}$ correspond to the identity of the faulty antenna elements. 

\begin{figure}[t!]
    \centering
    \includegraphics[width=270pt]{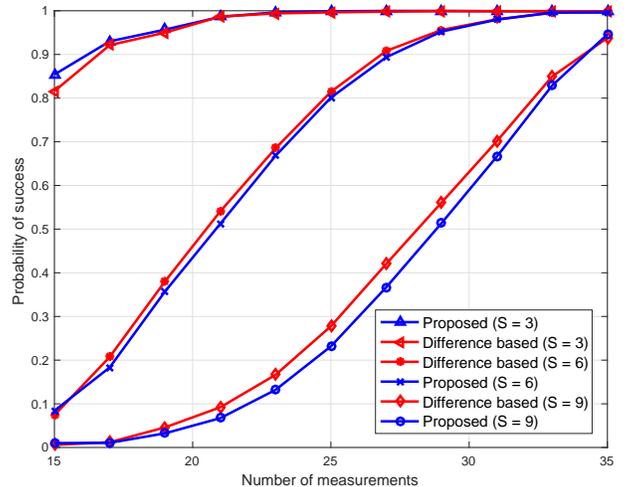} 
    \caption{Probability of successful fault detection versus the number of diagnostic measurements for different number of faults $S$ and single channel path. Probability of successful detection increases with the number of measurements.} \label{fig1}
  \end{figure}

As outlined, the proposed technique only requires knowledge of incident AoAs, and not the full channel knowledge, to recover the identity of the faulty antenna elements. This partial channel requirement reduces the implementation complexity of remote array diagnostic techniques. This, however, comes at the expense of additional complexity in the design of the receive antenna weights.

  \begin{figure}[t!]
    \centering
    \includegraphics[width=270pt]{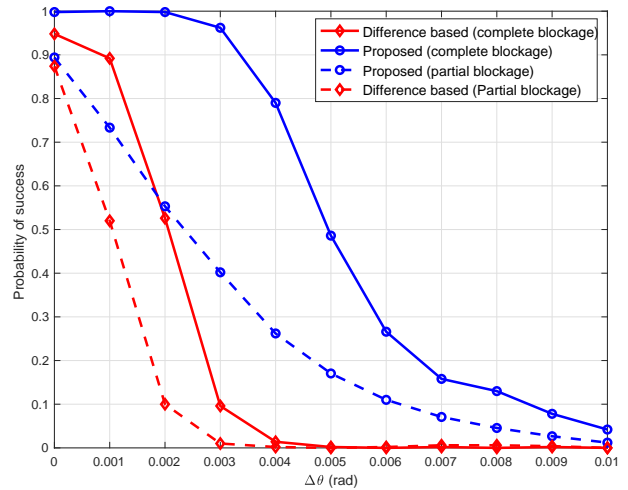} 
    \caption{Probability of successful fault detection versus AoA estimation error $\Delta \theta$ for $S=6$ random faults, $M=35$ measurements and single channel path. Proposed technique is less sensitive to estimation errors when compared to the difference based technique which requires full channel knowledge.} \label{fig2}
  \end{figure}
  
%\label{y2}
%blockages and receives diagnostic symbols from random AoAs, i.e. $\theta_{\ell} \in [0:2\pi]$
\section{Numerical Results and Discussions}
In this section, we conduct numerical simulations to evaluate the efficacy of the proposed technique. We consider a receiver with a uniform linear array with half wavelength separation and $S$ faulty antenna elements. We adopt the blockage and channel model presented in Section II. To generate complete antenna element blockages, randomly selected $S$ diagonal entries in the  blockage matrix $\mathbf{B}$ in (\ref{y2}) are set to zero. To generate partial blockages, $S$ diagonal  entries in the blockage matrix $\mathbf{B}$ are set to have a random phase shift and amplitude (see (\ref{efbp1})). We adopt the probability of success $\text{P}_\text{success} $, i.e. the probability that all faulty antennas are detected, as a performance measure to quantify the error in detecting the faulty antenna locations. This probability is defined  as
\[
\text{P}_\text{success} = \text{Pr} ({\mathcal{I}_S = \mathcal{\hat{I}}}_S),
\]
where the entries of the set $\mathcal{I}_S$ represent the \textit{true} identities of the faulty antennas and the entries of the set $\hat{\mathcal{I}}_S$ represent the identities of the \textit{detected} faulty antennas. For benchmarking purposes, we compare the  probability of success achieved by the proposed technique with the probability of success achieved by the difference based technique proposed in \cite{m0}. In all simulations, we set $N$ = 128 antennas and consider both single and multi-path channel cases.

For the single path case, the transmitting probe is situated so as to correspond to one of the $N=128$ quantized AoAs in the DFT matrix $\mathbf{A}$ (see (\ref{yp2})). The receive antenna weights are selected from the columns of $\mathbf{A}$ that do not correspond to the receiver's quantized AoA. The performance of the proposed technique for this scenario is illustrated in Fig. \ref{fig1} and Fig. \ref{fig2}. 

In Fig. \ref{fig1}, we plot the probability of success versus the number of measurements (or diagnosis time) for different number of antenna faults. Fig. \ref{fig1} shows that the proposed technique is able to successfully detect antenna faults without additional diagnostic measurements when compared to difference based techniques. This is achieved without the need for prior knowledge of the receiver's channel gain (or path-loss).

%This is achieved without the need for the receive channel's gain and additional diagnostic measurements when compared to difference based techniques as shown in Fig. \ref{fig1}. 

%To draw some insights into the effect of the channel estimation error on the performance of the propsed technique, 
In Fig. \ref{fig2}, we study the effect of AoA estimation errors on the performance of the proposed technique. Specifically, we plot the probability of success versus the AoA estimation error when the array is subjected to both complete and partial blockages. In the presence of partial blockages, Fig. \ref{fig2} shows that both the proposed and difference based techniques experience a slight loss in the detection performance when compared to complete blockages. This is mainly due to the fact that the magnitude of the errors in the error vector $\mathbf{h}_\text{e}$ in (\ref{yp32}) are smaller when compared to complete blockages. This effectively reduces the detection capability in the presence of noise when compared to the case of full blockages, which results in higher error magnitudes. Fig. \ref{fig2} also shows that both the proposed and difference based technique are sensitive to AoA estimations errors. Nonetheless, the proposed technique is superior in the sense it can tolerate significantly higher AoA errors. 

   \begin{figure}[t!]
    \centering
    \includegraphics[width=270pt]{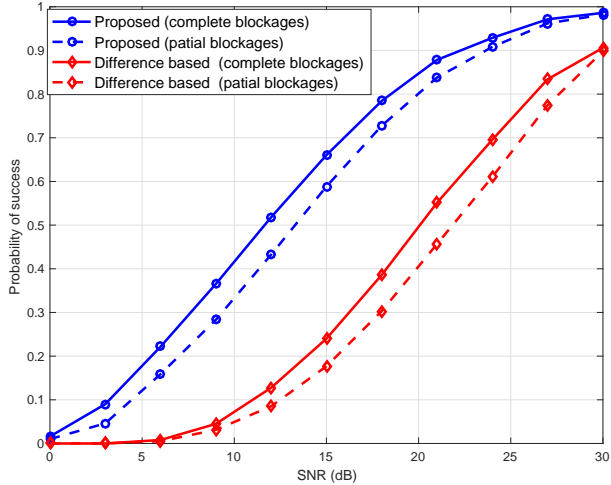} 
      \caption{Probability of successful fault detection versus the receive SNR for $S=6$ faults, $L=3$ channel paths and $M=35$ diagnostic measurements. The proposed technique is robust against system noise compared to the difference based technique.} \label{fig3}
 %   \caption{Probability of success versus AoA estimation errors in a uniform linear array subjected to $S = 6$ blockages. Proposed technique is less sensitive to estimation errors when compared to difference-based techniques that require full CSI. L = 3 paths, SNR = 30 dB, M = 35 } \label{fig3}
  \end{figure}

    \begin{figure}[t!]
    \centering
    \includegraphics[width=270pt]{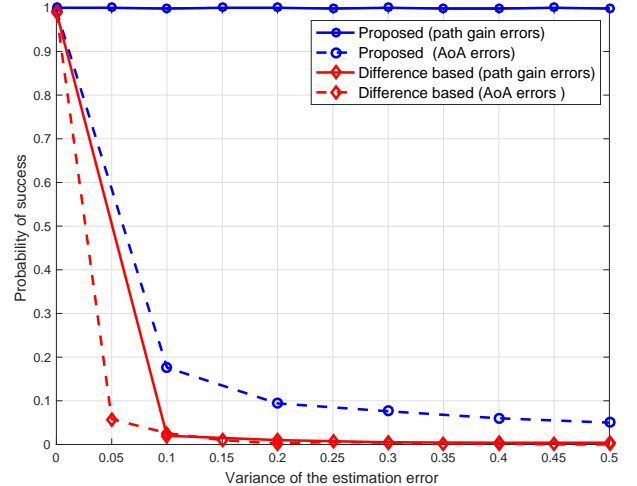} 
        \caption{Probability of successful fault detection versus the variance of the channel estimation error for $S=6$ faults, $L=3$ channel paths, SNR = 30dB and $M=45$ diagnostic measurements. The proposed technique is agnostic to path gain errors.} \label{fig4}
  \end{figure}

In Fig. \ref{fig3} and  Fig. \ref{fig4} we study the performance of the proposed technique in the presence of multi-path, complete and partial blockages, and non-quantized AoAs. Specifically,  each random path corresponds to a random AoA and complex gain. Based on knowledge of all AoAs, and using Householder transformation, the measurement matrix $\mathbf{W}$ is designed to result in a beam pattern that is orthogonal to the receiver's AoAs (see (\ref{HH})-(\ref{yp32})). In Fig. \ref{fig3} we plot the success probability versus the receive signal-to-noise ratio (SNR) in the presence of complete and partial blockages. Fig. \ref{fig3} shows that both the proposed and the difference technique require high SNR to successfully detect antenna faults, and the proposed technique is superior in the sense that it is less sensitive to the system noise. Note that noise affects both the path gains and the AoAs. As the proposed technique is mainly sensitive to AoAs errors, and not path gain errors, it experiences less performance degradtaion when compared to the difference technique. 

To draw some insights into the effect of channel estimation errors on the performance of the proposed technique, we plot the probability of success versus the variance of the path gain and AoA estimation errors at the receiver in Fig. \ref{fig4}. The estimation errors are assumed to be Gaussian distributed with zero mean and variance as indicated in Fig. \ref{fig4}. Interestingly, and as evident from Fig. \ref{fig4}, the proposed technique permits successful antenna fault detection irrespective of the path gain noise error magnitude. This performance gain is attributed to the fact that the proposed technique does not require knowledge of channel gain for fault detection. Hence channel gain estimation errors do not affect the performance of the proposed technique. Nonetheless, the proposed technique is shown in Fig. \ref{fig4} to be sensitive to AoA mismatch. As the mismatch increases, the orthogonality between the designed beamforming weights and the true channel response diminishes. This increases the noise at the receiver.    Fig. \ref{fig4} shows that the probability of success for the difference based technique deteriorates drsticlly with slight path gain or AoA mismatch. The reason for this is that any mismatch between the generated channel and the true channel would destroy the sparsity property of the difference channel response $\mathbf{h}_\text{d}$, and hence sparse recovery would not be possible in this case.

%In the presence of channel estimation errors, which could arise in the presence of antenna faults, Fig. \ref{fig4} shows that the proposed technique achieves better detection performance when compared to  the difference technique. This makes it a better choice for remote array diagnosis

\section{Conclusion}
In this paper, we proposed a novel array diagnosis technique for mmWave systems with large antenna arrays. The proposed technique is able to identify the locations of antenna faults with only partial channel knowledge. For both single path and multipath cases, the proposed technique is shown to be less sensitive to channel estimation errors when compared to the widely adopted difference based technique. This improvement comes at the expense of additional complexity in the design of the receive beamforming weights. Due to its robustness against channel estimation errors, the proposed technique can be deployed to perform real-time array diagnosis. Future work will focus on array diagnosis in the absence of any channel knowledge and on applying this technique on a practical set-up.

\section*{Acknowledgment}
This material is based upon work supported in part by the Sacramento State Research and Creative Activity Faculty Awards Program.

\end{document}